\documentclass[preprint]{aastex}
\usepackage{emulateapj5}
\usepackage{apjfonts}
\usepackage{amssymb, amsmath}
\usepackage{natbib}
\usepackage{graphicx}
\usepackage{CJK}

\input psfig.tex

\newdimen\digitwidth    
\setbox1=\hbox{0}       
\digitwidth=\wd1        
\catcode`"=\active      

\def"{\kern\digitwidth}

\def\aj{{AJ}}

\def\apj{{ApJ}}
\def\apjs{{ApJS}}
\def\asec{$^{\prime\prime}$}
\def\deg{$^{\circ}$}

\def\etal{{et al.\  }}

\def\lax{{$\mathrel{\hbox{\rlap{\hbox{\lower4pt\hbox{$\sim$}}}\hbox{$<$}}}$}}
\def\gax{{$\mathrel{\hbox{\rlap{\hbox{\lower4pt\hbox{$\sim$}}}\hbox{$>$}}}$}}
\def\simlt{\lower.5ex\hbox{$\; \buildrel < \over \sim \;$}}
\def\simgt{\lower.5ex\hbox{$\; \buildrel > \over \sim \;$}}

\def\mnras{{MNRAS}}

\def\galfit{{\tt GALFIT }}

\slugcomment{Submitted to {\it The Astrophysical Journal Letters.}}
\shorttitle{Two-phase Formation of Elliptical Galaxies}
\shortauthors{HUANG et al. }

\begin{document}

\begin{CJK*}{UTF8}{gbsn}

\title{Fossil Evidence for the Two-phase Formation of Elliptical Galaxies}

\author{Song Huang (黄崧)\altaffilmark{1,2,3} Luis C. Ho\altaffilmark{2}, 
Chien Y. Peng\altaffilmark{4}, Zhao-Yu Li (李兆聿)\altaffilmark{5}, and 
Aaron J. Barth\altaffilmark{6} }
\date{}                                          

\altaffiltext{1}{School of Astronomy and Space Science, Nanjing University,
Nanjing 210093, China}

\altaffiltext{2}{The Observatories of the Carnegie Institution for Science, 
813 Santa Barbara Street, Pasadena, CA 91101, USA}

\altaffiltext{3}{Key Laboratory of Modern Astronomy and Astrophysics, Nanjing
University, Nanjing 210093, China}

\altaffiltext{4}{Giant Magellan Telescope Organization, 251 South Lake Avenue,
Suite 300, Pasadena, CA 91101, USA}

\altaffiltext{5}{Key Laboratory for Research in Galaxies and Cosmology,
Shanghai Astronomical Observatory, Chinese Academy of Sciences,
80 Nandan Road, Shanghai 200030, China}

\altaffiltext{6}{Department of Physics and Astronomy, 4129 Frederick Reines 
Hall, University of California, Irvine, CA 92697-4575, USA}

\begin{abstract}

Massive early-type galaxies have undergone dramatic structural evolution over
the last 10 Gyr.  A companion paper shows that nearby elliptical
galaxies with $M_{\ast}\ge1.3\times10^{11} M_{\odot}$ generically contain 
three photometric subcomponents: a compact inner component with effective radius
$R_e$ \lax\ 1 kpc, an intermediate-scale middle component with $R_e \approx 2.5$ 
kpc, and an extended outer envelope with $R_e \approx 10$ kpc.  Here we attempt 
to relate these substructures with the properties of early-type galaxies 
observed at higher redshifts.  We find that a hypothetical structure formed from 
combining the inner plus the middle components of local ellipticals follows a
strikingly tight stellar mass-size relation, one that resembles the distribution 
of early-type galaxies at $z\approx1$.  Outside of the central kpc, the median 
stellar mass surface density profiles of this composite structure agree closest
with those of massive galaxies that have similar cumulative number density at 
$1.5<z<2.0$ within the uncertainty.  We propose that the central substructures 
in nearby ellipticals are the evolutionary descendants of the ``red nuggets'' 
formed under highly dissipative (``wet'') conditions at high redshifts, as 
envisioned in the initial stages of the two-phase formation scenario recently 
advocated for massive galaxies.  Subsequent accretion, plausibly through 
dissipationless (``dry'') minor mergers, builds the outer regions of the 
galaxy identified as the outer envelope in our decomposition.  The large 
scatter exhibited by this component on the stellar mass-size plane testifies
to the stochastic nature of the accretion events.
\end{abstract}


\maketitle

\section{Introduction}

Recent observations have established that high-redshift early-type galaxies 
(ETGs) are more compact (Daddi \etal 2005; Trujillo \etal 2006; Damjanov et 
al. 2011) and have higher stellar velocity dispersions (Cappellari \etal 2009;
Onodera \etal 2009) than nearby ETGs of the same stellar mass. Since 
$z \approx 2.5$, ``red nuggets'' on average have doubled in stellar mass and 
increased their size by a factor of $3-4$ (Buitrago \etal 2008; van Dokkum et 
al. 2010; Papovich \etal 2012; Szomoru \etal 2012).  Some appear to have a 
disk-like morphology (van der Wel \etal 2011), and they are always bluer in 
the outskirts (Gargiulo \etal 2012).  The accumulated evidence suggests 
that massive ETGs build up inside-out through non-dissipational processes 
(Bezanson \etal 2009; van Dokkum \etal 2010). 
 
These developments severely challenge classical models of elliptical (E) galaxy 
formation such as monolithic collapse (Larson 1975) and binary major mergers 
(Toomre \& Toomre 1972; Negroponte \& White 1983).  Instead, the current 
evidence favors a ``two-phase'' scenario (Oser \etal 2010; Johansson \etal 
2012). Intense dissipational processes such as cold accretion (Dekel \etal 
2009) or gas-rich mergers rapidly build up an initially compact progenitor. 
After star formation is quenched, a second phase of slower, more protracted 
evolution is dominated by non-dissipational processes such as dry, minor
mergers. 

For such a scenario to work, the rate of minor mergers (Bluck et al.\
2012; Newman et al.\ 2012) has to be consistent with the prevalence of
faint companion galaxies at high redshifts.  Perhaps other mechanisms may
also be important, such as major merger (Bernardi et al.\ 2011; Prieto et al. 
2013) or AGN-induced expansion (Fan et al.\ 2008). Given that large 
uncertainties remain in identifying the relevant physics and quantifying their 
detailed balance, a key sanity check is to see whether one can separate out the
fossil bodies of compact $z > 1.5$ galaxies at the core of nearby galaxies.
This separation is successful only when several clear and solid predictions, 
especially the similarity between the core components and the high-$z$ compact 
progenitors in mass versus size relation and detailed mass profile, are met.  
These stringent predictions are the basis of this study to verify or falsify.
We assume a $\Lambda$CDM cosmology with 
$\Omega_m=0.27$, 
$\Omega_{\Lambda}=0.73$, and $H_0=73$ kms$^{-1}$ Mpc$^{-1}$.

\section{Observational Material}

This work uses the three-component models of nearby Es from
the Carnegie-Irvine Galaxy Survey (CGS; Ho \etal 2011), a systematic
study of 605 bright ($B_T < 12.9$ mag) southern ($\delta<0$\deg) galaxies. The
currently completed photometric part of CGS includes \emph{BVRI} images
obtained using the 100-inch du Pont telescope at Las Campanas Observatory,
using a 8\farcm9$\times$8\farcm9 detector with a pixel scale of $0\farcs259$,
under $\sim$1\asec\ seeing conditions. More details on the observations and
data reductions can be found in Ho \etal (2011) and Li \etal (2011).

\vskip 0.15cm
\begin{figure*}[htb]
\centerline{\psfig{file=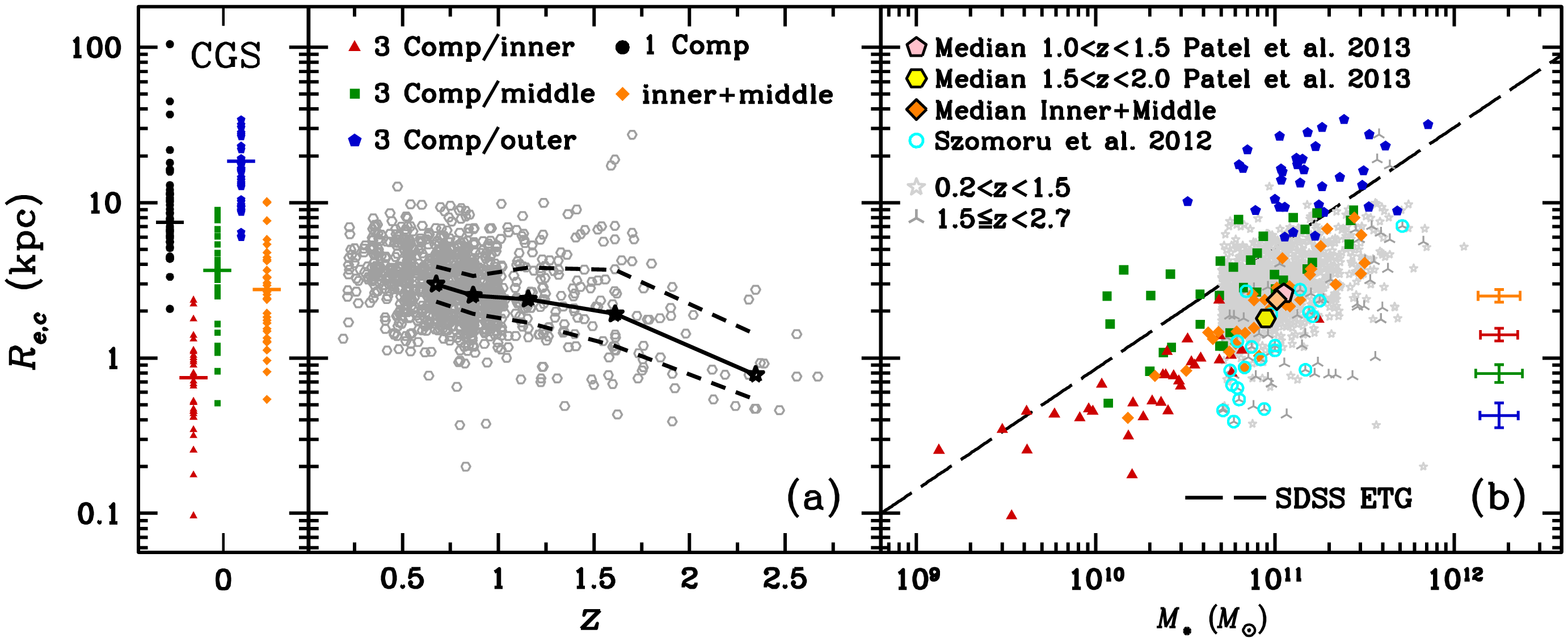,width=18.5cm,angle=270}}
\figcaption[fig1.eps]{
The redshift evolution of size and its relation with stellar mass for nearby 
Es and high-$z$ ETGs. The left side of panel (a) summarizes the size 
distributions of single-component models (black circles) and the inner (red 
triangles), middle (green squares), and outer (blue pentagons) components of 
the three-component models of CGS Es. The median of each distribution is 
marked by a 
horizontal bar. The right side illustrates the redshift evolution for high-$z$ 
ETGs with $M_{\ast} \ge 10^{10.7}\,M_{\odot}$ (grey open circles).  The solid
and dashed lines give the running median and its associated first and third 
quartile through five redshift bins ($0.50 \le z<0.75$; $0.75\le z<1.0$; 
$1.0\le  z<1.4$; $1.4\le  z<2.0$; $2.0\le  z\le 2.7$). Panel (b) shows the 
mass-size relation for high-$z$ ETGs, separated into those with $z < 1.5$ 
(grey stars) and $z \ge 1.5$ (grey skeletal triangles).  The 21 massive ETGs at
$1.5< z<2.5$ from Szomoru \etal (2012) are marked with cyan circles.  
Each of the three subcomponents of the CGS sample is plotted separately, as 
well as for the inner$+$middle components combined (orange diamonds).  Also 
highlighted are the median properties for the inner$+$middle components 
of the CGS Es and for the massive galaxies at $1.0<z<1.5$ and $1.5<z<2.0$ 
selected by their cumulative number density from Patel \etal (2013).
Typical error bars for the subcomponents are 
shown on the lower-right corner.  The best-fit relation of Guo et 
al. (2009) for local SDSS ETGs is plotted with a dashed line.  For consistency 
with the high-$z$ measurements, the effective radii used here are the so-called 
mean, or circularized, radii, such that $R_{e,c}=R_e\, \sqrt{b/a}$, where $b/a$
is the axis ratio.  
\label{fig1}}
\end{figure*}
\vskip 0.1cm

Huang \etal (2013) apply \galfit (Peng \etal 2010) to perform a 
detailed two-dimensional analysis of the $V$-band images of 94 CGS Es.  Their 
decomposition reveals that, contrary to popular perception, the global light 
distribution of most nearby Es have more complicated structures than can be 
described by a single S\'ersic (1968) profile.  Instead, 70 out of the 94 Es 
can be fit with three distinct subcomponents: (1) an inner, compact center with 
effective radius $R_e$ \lax\ 1 kpc comprising a small fraction of the light 
($f \approx 0.1-0.15$); (2) an intermediate-scale, middle structure with $R_e 
\approx 2.5$ kpc and $f \approx 0.2-0.25$; and (3) an outer, dominant, 
extended, moderately flattened envelope with $R_e \approx 10$ kpc and $f 
\approx 0.6$.  

The image decomposition was performed without ascribing any physical 
significance to the substructure.  Nevertheless, the fact that the individual
subcomponents define distinct sequences on the size-luminosity plane, and that
three-component fits often seemed naturally preferred, hint at the possibility 
that the subcomponents reflect underlying physical reality.  Here we compare
some key properties of the three-component models of local Es with observations 
of high-$z$ ETGs to shed new light on the structural evolution of massive 
galaxies.

For this purpose, we compiled stellar masses and effective radii for 1323
high-$z$ ETGs with stellar masses $M_{\ast} \ge 10^{10.7}\,M_{\odot}$, 
as follows: 352 ($0.2<z<2.7$) from Damjanov \etal (2011); 910 ($0.2<z<1.2$) 
from COSMOS (Scoville \etal 2007; using morphological classification from 
Scarlata \etal 2007 and Tasca \etal 2009); 32 ($1.5<z<3$, choosing S\'ersic 
indices $n>2$) from the GOODS-NICMOS survey (Conselice \etal 2011); 8 
($z\approx 1.6$, $n>2$) from Papovich \etal (2012); and 21 ($1.5<z<2.5$) from 
CANDELS (Grogin \etal 2011, Koekemoer \etal 2011), as analyzed by 
Szomoru \etal (2012).  Although the sample is in many respects heterogeneous 
(in terms of observed bandpass, selection criteria, modeling method), the 
general trends we explore in this paper should not be strongly affected, as
demonstrated by Damjanov \etal (2011).

\section{Linking Nearby and High-$z$ Early-type Galaxies}

Figure~1(a) illustrates the redshift evolution of galaxy size for massive 
ETGs. Although the scatter is substantial, the size increase with decreasing 
$z$ is confirmed (e.g., Damjanov \etal 2011; Cimatti \etal 2012). The median 
circularized effective radius increases from $0.8$ kpc at $2.0\le z\le 2.7$, to 
$R_{e,c} \approx1.9$ kpc at $1.4\le z<2.0$, to $R_{e,c}\approx3$ kpc at 
$0.50\le z<0.75$. By $z=0$, Es have reached a median $R_{e,c}=7.5$ kpc 
according to the single-component fits applied to the subsample of 35 massive 
CGS Es that satisfy $M_{\ast}$ \gax\ $10^{11}\, M_{\odot}$ (solid black 
points). The CGS mass cut roughly corresponds to twice the mass 
limit of the high-$z$ sample.

How are the subcomponents of nearby Es related to the high-$z$ systems? Judging 
by the distribution of sizes for the subcomponents shown on the left side of 
Figure~1(a), the typical effective radius of the inner component (median 
$R_{e,c} = 0.9$ kpc) does overlap substantially with that of high-$z$ red 
nuggets; however, their stellar masses, only $\sim$10\%--15\% of the total, 
are about a factor of 3--5 lower than the high-$z$ systems (Figure~1b).
The middle component, with median $R_{e,c} = 3.7$ kpc, matches well the sizes 
and masses of ETGs at $z\approx 0.5$, but they are still on average more 
extended than those at $z>1.0$.  On the other hand, with median $R_{e,c} = 17.5$
kpc, the outer component of local Es is clearly larger than the globally 
averaged size of most ETGs at any redshift. 

\vskip 0.2cm
\begin{figure*}[htb]
\centerline{\psfig{file=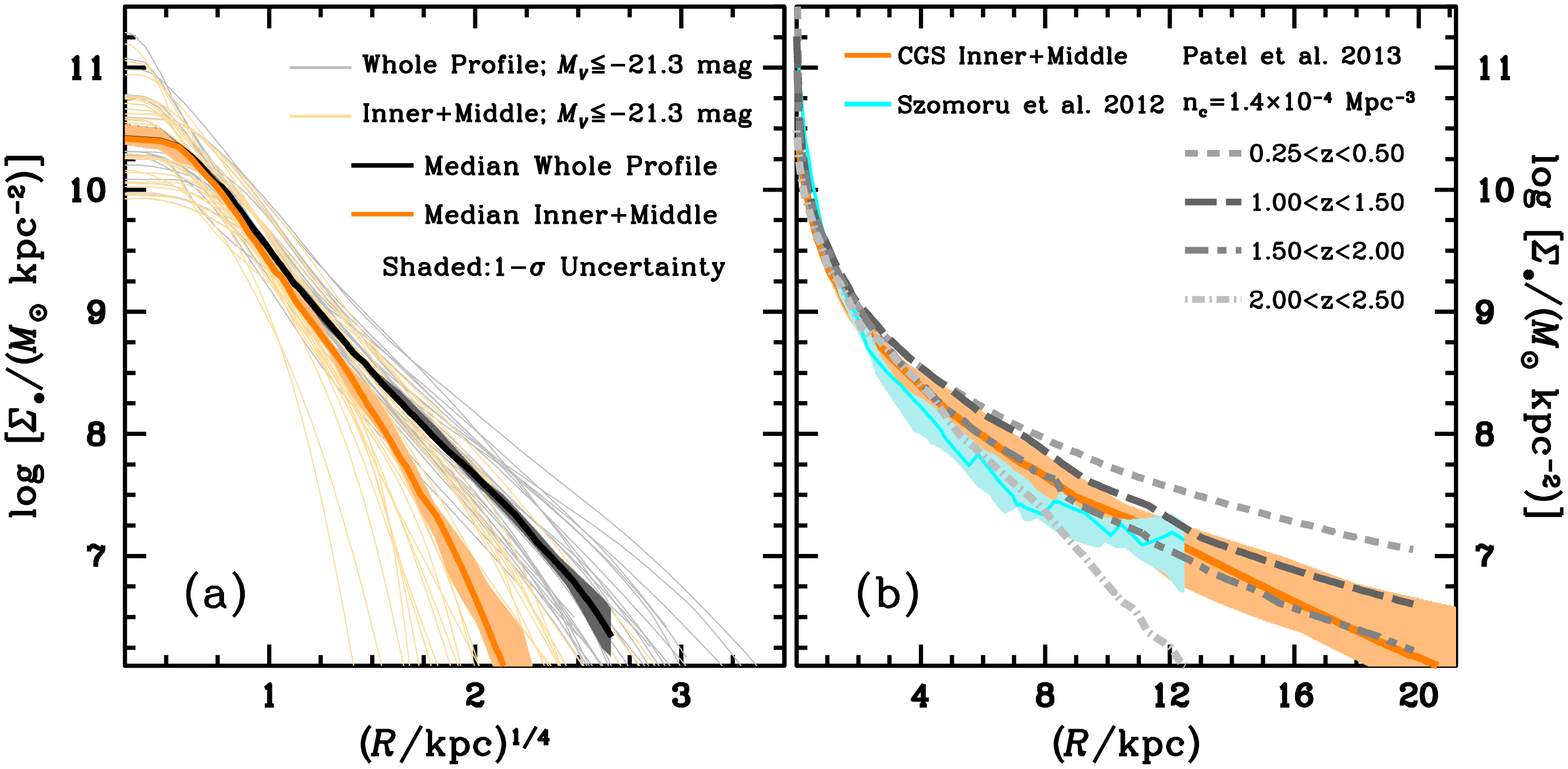,width=18.50cm,angle=270}}
\figcaption[fig2.eps]{
(a) Mass surface density profiles for CGS Es with $M_V \leq -21.3$ mag.  The 
individual profiles for the entire galaxy and for the combined inner$+$middle
components are plotted in light grey and orange, respectively; their 
corresponding median profiles are plotted as darker solid lines, with the 
dashed lines denoting the associated 1~$\sigma$ uncertainty range of the median 
value (determined from bootstrap resampling).  (b) Comparison of the median 
profile of the inner$+$middle components for CGS Es (orange solid line; dashed 
lines 1~$\sigma$ uncertainty) with the median profiles for massive galaxies at 
different redshift bins selected by their cumulative number density 
($n_c = 1.4\times10^{-4}$\ {\rm Mpc}$^{-3}$; Patel \etal 2013). The median 
profile (and associated uncertainty) for the 21 massive ETGs at $1.5<z<2.5$ from 
Szomoru \etal (2012) is also shown.  Note that the radial scale is 
plotted as $R^{1/4}$ in (a) but as linear in $R$ in (b) to emphasize the outer
regions.
\label{fig2}}
\end{figure*}
\vskip 0.3cm

The mass-size relation offers another vantage point.  As described in
Huang \etal (2013), single-component fits to the CGS Es produce sizes that 
follow well the mass-size relation of local ETGs selected from the Sloan Digital 
Sky Survey (SDSS; Guo \etal 2009, dashed line in Figure~1b).
As expected, high-$z$ ETGs, especially those with $z\ge 1.5$, fall 
systematically below the local relation.  At fixed stellar mass, high-$z$
systems are a factor $\sim 4-5$ more compact than their $z = 0$ counterparts.
While the inner component of the CGS galaxies is generally not massive enough 
to directly map onto high-$z$ objects, the middle component can be viewed as a 
slightly modified version of the compact quiescent galaxies.  The mass-size 
relation for the middle component runs roughly parallel to, but is offset 
slightly below, that of local ETGs.  It lies approximately in between the locus 
of high-$z$ and low-$z$ points.  In dramatic contrast with the inner and middle 
components, the outer envelope displays significantly larger scatter on the 
$R_e-M_{\ast}$ plane.  While some of the scatter undoubtedly arises from 
measurement uncertainty, which, as indicated at the bottom-right corner of 
Figure~1(b), is considerable for extended, low-surface brightness features, we 
believe that most of the scatter is intrinsic. 

As discussed in Huang \etal (2013), the central core may have been reshaped more
recently (Hopkins et al. 2009a, b, c) than the overall compact stellar body of
interest, thus requiring a separate component to model well in order to
sensitively extract the lower lying outer component. Regardless of the exact
process that shaped the small central structure, it is a small perturbation to
the overall structure of interest, hence we group the innermost two components
(designated henceforth as "inner$+$middle") into a single entity.  To measure 
the overall size, mass, and profile of the composite, we generated model images
from best fit parameters that summed the innermost subcomponents, accounting for
the PSF.  Then, we fit a single-component model to extract the effective radius 
of this composite structure.  As the inner component has higher density than the
middle one, we intuitively expect the effective radius of the combined structure
to be smaller than that of the middle component alone. This is indeed the case,
as Figure~1(b) shows.  Surprisingly, the composite inner$+$middle structure 
(orange diamonds) not only forms a very tight mass-size relation with a slope 
similar to that of Guo \etal (2009) for SDSS ETGs, but it also lies on top of 
the distribution of points for most high-$z$ ETGs, especially those with 
$z<1.5$.  

As advocated by van Dokkum \etal (2010), galaxies selected at constant 
cumulative number density provide a more physical connection between 
progenitors and descendants than those selected above a constant stellar mass.
We therefore compare our observations with the sample of Patel \etal (2013), 
which is selected at a cumulative number density ($n_c = 1.4 \times 10^{-4}$\ 
${\rm Mpc}^{-3}$) that corresponds to massive ETGs ($M_{\ast} \approx 10^{11.2} 
\, M_{\odot}$) similar to the CGS sources at $z \approx 0$.  Figure~1(b) shows 
that the median $M_{\ast}$ and $R_{\it e}$ of our inner$+$middle components 
agree quite well with the properties of galaxies at $1.0<z<1.5$ from Patel et 
al. (2013). 
Apart from global quantities such as mass and effective radius, the surface
density profiles themselves provide a more direct comparison of the
structural connection between high-$z$ ETGs and local Es.  For this purpose,
we convert the surface brightness profiles of the inner$+$middle components,
after removing the point-spread function smearing, into stellar mass density
profiles using the mass-to-light ratio adopted in Huang \etal (2013).  Their
median profile is shown in Figure~2(a), and in Figure~2(b) we compare it
to the median profiles of the cumulative number density-selected galaxies in
different redshift bins from Patel \etal (2013).  The median profile of the
inner$+$middle components of local Es matches quite closely the profiles of
massive galaxies at $1.5<z<2.0$ over most of their radial extent.  Unlike the 
median $M_{\ast}$ and $R_{\it e}$, the median profile of $1.0<z<1.5$ massive 
galaxies only marginally close to the profile of inner$+$middle component.  This
seemingly difference relates to the higher density of high-$z$ objects at 
$R<1.0$ kpc. It is confirmed in the median density profile of the massive
ETGs at $1.5<z<2.5$ from Szomoru \etal (2012), which show even higher central
densities than the sample from Patel \etal at comparable redshifts.  This 
apparent discrepancy may be due to different selection criteria employed in the 
two studies. 

The two-phase formation scenario of ETGs (Oser \etal 2010; Johansson et 
al. 2012) offers an appealing framework to interpret the above trends and to 
connect the distant and local populations of massive galaxies.  Dissipative 
processes (``in situ'' star formation, in the language of Oser \etal 2010) 
connected with early gas-rich events naturally lead to high central densities.  
We suggest that the central, more compact structures of nearby Es---namely 
the ``inner'' or ``middle'' components, possibly both---are the natural 
outcomes of this initial, dissipative phase.  We propose that they are the 
direct remnants of the high-$z$ massive quiescent galaxies. The combined 
luminosity (mass) fraction of the inner$+$middle components, $f \approx 
0.3-0.4$ are roughly comparable to the factor of $\sim 2$ increase necessary 
to grow the high-$z$ objects, if they are the progenitors of local Es.  Their 
overall physical scales are also consistent with their originally compact 
nature.  In detail, the current evidence suggests that the high-$z$ objects 
tend to have somewhat higher central densities than their local remnants. 
This is an important constraint that future theoretical models should try
to reproduce.

\vskip 0.15cm
\begin{figure*}[t]
\centerline{\psfig{file=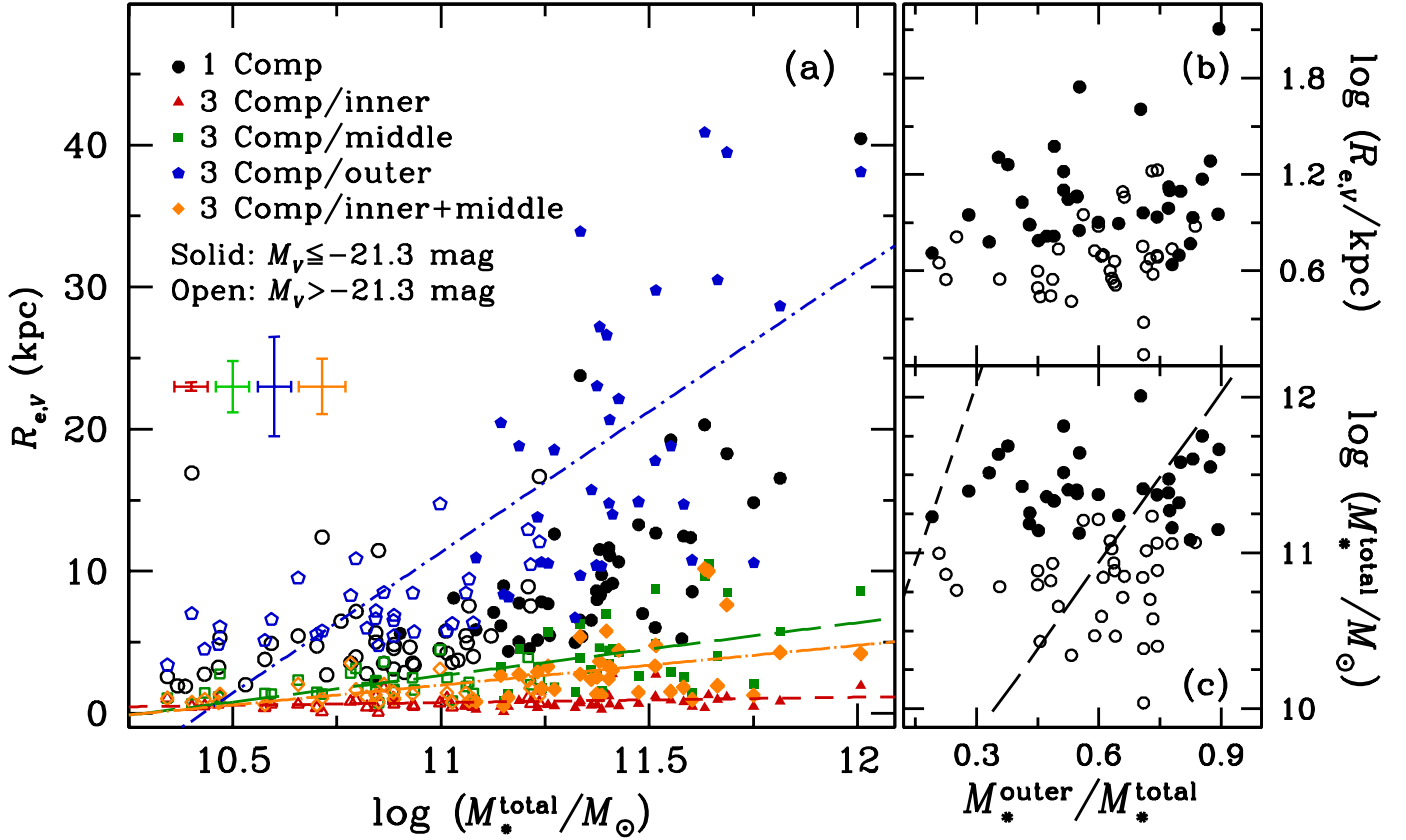,width=18.5cm}}
\figcaption[fig3.eps]{
(a) Correlations between $R_e$ of different components and total stellar mass
$M_{\ast}^{\rm total}$.  The symbols are the same as in Figure~1(a); the
linear correlations from least-squares fitting are shown using the same color
scheme as the points.  The two right panels show the dependence between the
mass fraction of the outer component and $R_e$ from single-component models (b)
and $M_{\ast}^{\rm total}$ (c).  The two lines in panel (c) represent the
relation between accreted mass fraction (which we approximate with
$M_{\ast}^{\rm outer}$/$M_{\ast}^{\rm total}$) and $M_{\ast}^{\rm total}$ from
the simulations of Oser \etal (2010; long-dashed line) and Lackner \etal
(2012; short dashed line).  Note that for the purposes of this comparison we
make use of the full sample 70 CGS Es with robust multi-component fits, not
just the high-luminosity subsample.
\label{fig3}}
\end{figure*}
\vskip 0.1cm

We equate the extended outer envelope with material accreted through 
dissipationless minor mergers (``ex situ'' stars; Oser \etal 2010).  The large 
intrinsic scatter on the $R_e-M_{\ast}$ plane is likely related to the 
stochastic nature by which the outer envelope is expected to have accumulated.  
The minor merger hypothesis has previously been questioned on the basis of
the apparently small observed scatter of the mass-size relation of nearby Es 
(Nipoti \etal  2012).  For those two observations to be consistent, dry mergers
would have to preferentially add material to large radii, realized through many
stochastic events spread over a protracted period.  We propose that this history 
is represented by the extended, outer component identified through our 
photometric decomposition.

We end by briefly comparing our decomposition results with predictions from
recent numerical simulations for the formation of massive ETGs.
After separating the final stellar system into its in situ
(dissipative) and ex situ (accreted) constituents in their simulations,
Oser \etal (2010) show that, in comparison to the in situ material, the
typical size of the accreted component should correlate with the total
stellar mass of the system.  Furthermore, the stellar mass fraction of the
accreted component may also correlate with the total stellar mass and
effective radius of the whole system (Oser \etal 2010; Johansson \etal
2012; Lackner \etal 2012).  However, their numerical simulations produce a
wide range of possible outcomes: the predictions for the variation of accreted 
fraction with total stellar mass in Oser \etal (2010) disagree strongly with 
those from the work of Lackner \etal (2012).

We compare our analysis with the simulations to look for correlations in
the accreted components.  We assume that the outer and
inner+middle components of our multi-component fits approximate the 
``accreted'' and ``in situ'' material in the theoretical
models.  Figure 3(a) shows that the effective radius of the outer component
of nearby Es does indeed increase strongly with the total stellar mass of
the system, much more so than either of the inner or middle components.
While the physical scales and the slope of the correlation are not the same
as predicted in the simulations of Oser \etal, there is qualitative agreement.
However, the comparison between the accreted fraction, approximated as
$M_{\ast}^{\rm outer}$/$M_{\ast}^{\rm total}$, and effective radius (Figures 3b)
and total stellar mass (Figures 3c) do not show much, if any, correlation.  
Apart from the uncertainties of the predictions of current simulations,
it is a difficult observational challenge to isolate the accreted component via 
photometric decomposition. The details of any decomposition are dependent on the
choice of specific parameterized model components, which are not unique choices.  
Also, further galaxy merging would have diluted and partly erased the signatures.

\section{Summary and Discussion}

Huang \etal (2013) show that the surface brightness distribution of nearby Es 
can best be described by three photometrically distinct substructures, 
consisting of a compact ($R_e$ \lax\ 1 kpc) inner component, an 
intermediate-scale ($R_e \approx 2.5$ kpc) middle component, and an extended 
($R_e \approx 10$ kpc) outer envelope.  Motivated by the recently proposed 
two-phase formation scenario for massive galaxies, here we argue that the 
inner and middle components identified in nearby Es jointly comprise a 
structure that can be considered the evolved, local counterparts of the 
high-$z$ compact massive galaxies. In this scenario, these compact structures 
derived from highly dissipative processes at high redshifts.  Dissipationless 
minor mergers dominate the late-time evolution of ETGs by
contributing to their dramatic size growth.  We identify the outer, extended 
component of nearby Es with this late accretion phase.

Several lines of investigation deserve further attention.  In light of the 
photometric subcomponents uncovered in nearby Es, it is of interest to ask 
whether such substructure already exists in high-$z$ ETGs, and, if so, whether 
their gradual build up can be detected as a function of time.  Current studies 
treat high-$z$ ETGs as single-component systems (e.g., Papovich \etal 2012; 
Szomoru \etal 2012), but the data may have sufficient quality to allow 
more complex analysis to reveal potential substructure.  
Of particular interest is the mechanism by which the ``core'' structure in 
massive Es (e.g. Lauer \etal 1995; Kormendy \etal 2009) arise.  If scouring by 
binary supermassive black holes is necessary to make cores (e.g., Hopkins 
\etal 2009b, c), then major mergers are implicated.  Similarly, if high-$z$ 
massive galaxies contain disky or fast-rotating structures (van der Wel \etal 
2011), then they must be destroyed, presumably also through major mergers, if 
they are to evolve into local massive Es.  How do major mergers
fit into the framework of the two-phase formation scenario?  


Beyond imprints on the photometric structural properties, the two-phase
formation scenario also implies differences in stellar population.  Recent
spectroscopic studies of the faint, outer regions of nearby Es already hint
that they have lower metallicity and possibly older age than the inner
parts (Coccato \etal 2010; Greene \etal 2012).  Photometry is vastly 
simpler and less expensive than spectroscopy, even if broad-band colors 
provide less robust constraints on stellar populations than spectra.
A forthcoming paper in will make use of the multi-band CGS data to extract
stellar population constraints from the color information of the subcomponents 
in nearby Es.

\acknowledgements
We thank the anonymous referee for helpful comments 
and S. Patel for kindly providing us the data from his work.
This work was supported by the Carnegie Institution for Science (LCH), 
the UC Irvine School of Physical Sciences (AJB), the China Scholarship 
Council (SH, Z-YL), and under the National Natural Science Foundation of 
China under grant 11133001 and 11273015 (SH).
SH thanks Prof. Q.-S. Gu for providing long-term support.


\end{CJK*}

\end{document}